\documentclass{emulateapj}

\shorttitle{Dancing twins}
\shortauthors{Tokovinin}

\begin{document}

\renewcommand{\topfraction}{1.0}
\renewcommand{\bottomfraction}{1.0}
\renewcommand{\textfraction}{0.0}

\title{Dancing twins: stellar hierarchies that formed sequentially?}


\author{Andrei Tokovinin}
\affil{Cerro Tololo Inter-American Observatory, Casilla 603, La Serena, Chile}
\email{atokovinin@ctio.noao.edu}

\begin{abstract}
This paper  attracts attention to  the class of resolved  triple stars
with moderate  ratios of inner and  outer periods (possibly  in a mean
motion  resonance)  and  nearly  circular,  mutually  aligned  orbits.
Moreover,  stars in  the inner  pair are  twins with  almost identical
masses, while the mass sum of the inner pair is comparable to the mass
of  the  outer  component.    Such  systems  could  be  formed  either
sequentially  (inside-out)  by   disk  fragmentation  with  subsequent
accretion and migration or  by a cascade hierarchical fragmentation of
a  rotating cloud.   Orbits  of  the outer  and  inner subsystems  are
computed  or  updated in  four  such  hierarchies: LHS~1070  (GJ~2005,
periods 77.6 and 17.25 years), HIP 9497 (80 and 14.4 years), HIP 25240
(1200 and 47.0 years), and HIP 78842 (131 and 10.5 years).
\end{abstract} 


\section{Introduction}
\label{sec:intro}

Interest  in the  architecture  and dynamics  of hierarchical  stellar
systems is stimulated by the desire to better understand their origin.
Low-mass hierarchies with outer separations less than $\sim$50 au show
a   tendency of  orbit alignment \citep{moments},  resembling in
this respect multi-planet  systems \citep{Fabrycky2014}, although
  they differ  from planets in other properties.  This suggests that such
hierarchies could have formed and evolved in a viscous accretion disk.
As the  gas from the  circumbinary disc is preferentially  accreted by
the secondary component, binaries  evolve toward equal masses (twins),
while their  orbits shrink in  size \citep{AL2000}.  This  scenario is
naturally  extended to explain  nearly co-planar  compact hierarchies:
while the  inner binary  shrinks, another companion  can be  formed by
fragmentation of  the circumbinary  disk, possibly destabilized  by an
accretion  burst.   The newly  formed  outer  companion overtakes  the
accretion and  grows until  its mass equals  the combined mass  of the
inner pair.  Thus,  formation of {\it double twins},  where both outer
and inner  mass ratios are close to  one, is a natural  outcome of the
sequential   inside-out   assembly   of   hierarchies.    The   triple
protostellar  system discovered by  \citet{Tobin2016} is  an excellent
illustration of  this scenario.   Its outer component  C is  the least
massive one,  yet it  accretes most of  the gas from  the circumbinary
disk and  will eventually catch  up in mass  with the inner  pair A,B,
unless the gas supply is  exhausted or dispersed earlier or its inward
migration destabilizes the system dynamically.

\begin{figure}
\epsscale{1.1}
\plotone{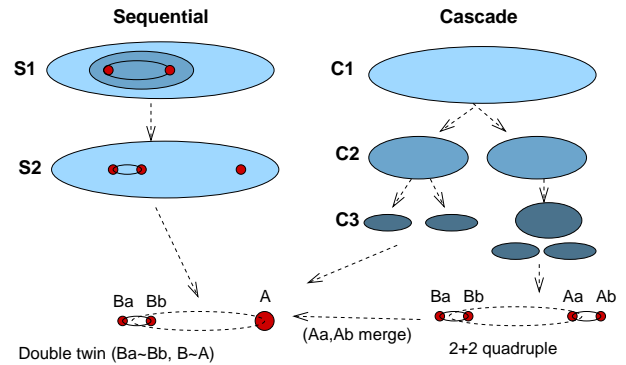}
\caption{Two  scenaria  leading to  the  formation  of   double-twin
  hierarchies with  approximately aligned orbits.   In the sequential,
  inside-out  formation (left),  the  inner binary  forms first  (S1),
  grows  in mass  and migrates,  until  the outer  companion is  formed
  (S2). The companion A can grow to the point when its mass equals the
  total mass of the inner  binary Ba,Bb, thus producing a double twin.
  In the  cascade (hierarchical) fragmentation  scenario (right), the rotating
  cloud fragments into approximately equal parts (C2), and then one or
  both fragments further split, forming one or two inner pairs (C3). A
  quadruple system (Ba,Bb and Aa,Ab) can later become a double twin if
  one of its inner pairs merges.
\label{fig:cartoon}  }
\end{figure}

However, the  sequential inside-out formation,  depicted schematically
in  Figure~\ref{fig:cartoon}  (left),  is  not the  only  possibility.
During  collapse  of pre-stellar  cores,  the  increasing density  and
decreasing Jeans  mass promote cascade fragmentation,  starting at the
largest  scales  and  proceeding  to  sub-fragmentation  into  smaller
parcels \citep{Cascade,Guszejnov2017}.  If  the mass is evenly divided
between  the  fragments,  cascade  fragmentation will  produce  a  2+2
quadruple system  with four nearly equal components,  or a double-twin
triple system if only one of  the outer parcels fragments or if one of
the    inner     pairs    merges    during     subsequent    evolution
(Figure~\ref{fig:cartoon},  right).  In this  scenario, masses  of the
emerging stars are set by the  masses of the fragments, while the size
of the  outer orbit is defined  by the angular momentum  of the parent
core.

Both formation scenarios probably occur. Massive 2+2 quadruple systems
of $\varepsilon$~Lyr  type with  wide outer orbits  \citep{Tok2008} match
the  expected outcome of  cascade fragmentation.   On the  other hand,
compact hierarchies composed  of low-mass stars are more  likely to be
formed  by   disc  fragmentation  sequentially,   inside-out.   Inward
migration  of  the  outer  companion  can disrupt  the  triple  system
dynamically. However,  before this happens, the  interacting outer and
inner orbits can be trapped in a mean motion resonance (MMR), as often
occurs  in  multi-planet  systems  \citep[e.g.][]{Lee2013}.   Orbital
motions in such hierarchies resemble a complex dance.

\begin{deluxetable*}{ccc lcc cc c }
\tabletypesize{\scriptsize}
\tablewidth{0pt}
\tablecaption{Overview of multiple systems
\label{tab:list} }
\tablehead{
\colhead{WDS} &
\colhead{HIP} & 
\colhead{$V$ (mag)} &
\colhead{Outer Name} & 
\colhead{$P$} & 
\colhead{$a$} &
\colhead{$M_1$}  &
\colhead{$M_2$}  &
\colhead{$\Phi$ (\degr)}  
  \\
          &
\colhead{$\pi$ (mas)} & 
\colhead{Sp.type} &
\colhead{Inner Name} & 
\colhead{(yr)} & 
\colhead{(\arcsec)} &
\colhead{ (${\cal M}_\odot$)} &
\colhead{ (${\cal M}_\odot$)} &
\colhead{$P_2/P_1$}  
}
\startdata
00247-2653 & \ldots & 15.4      & LEI 1 A,BC     &   77.6   & 1.528  &  0.12 & 0.15s & 1, 125 \\
           &  129.5   &  M7V    & LEI 1 B,C        &  17.25 & 0.460  & 0.077 & 0.070  & 4.50$\pm$0.03 \\
02022-2402 & 9497   & 7.88       &HDS 272 A,B  & 80   & 0.414 &  1.44 & 1.22s & 27, 27 \\
           & 16.09  & F6V        &TOK 41 Ba,Bb & 14.4 & 0.102 &  0.64 & 0.58  & 5.6$\pm$0.7    \\
05239-0052 & 25240  & 6.11       &WNC 2 A,BC   & 1200 & 3.716 &  2.73s & 2.62s & 17, 166 \\
           & 18.75  & F7V        &A 847 B,C    & 47    & 0.337 &  1.07  & 1.55  & $\sim$25         \\
16057-3252 & 78842  & 8.34       &SEE 264 A,B  & 131   & 0.807  & 0.96 & 1.50s & 14, 65 \\
           & 24.70  & K0V        &WSI 84 Ba,Bb & 10.5  & 0.128  & 0.75 & 0.75  & 12.6$\pm$0.2    
\enddata
\tablecomments{Explanation of columns: (1) WDS code \citep{WDS}; 
(2) Hipparcos number and trigonometric parallax from various sources; 
(3) combined visual magnitude and spectral type from Simbad; 
(4) discoverer codes; 
(5) orbital period; 
(6) semimajor axis;
(7) mass of the primary component;
(8) mass of the secondary component (s means mass sum);
(9) two values of mutual inclination $\Phi$ and the period ratio.
 }
\end{deluxetable*}

Detailed  observational characterization  of  masses and  orbits in  a
large  number  of hierarchies  will  help  to  decide which  formation
scenario  dominates and  will challenge  the theory  to  explain their
architecture.  In  this paper, the  observed motions in  four resolved
hierarchical  systems composed of  low-mass stars  are modeled  by the
inner  and outer  instantaneous Keplerian  orbits.  
Relative  orbit  orientations,  period   ratios,  and  masses  of  the
components are deduced, thus  enriching the still scarce observational
data  on  hierarchical  multiple  systems and  contributing  to  their
statistics,   like   the    previous   efforts   in   this   direction
\citep{planetary,trip}.   However, the  hierarchies  studied here  are
wider and slower  and do not have matching  radial velocity (RV) data;
their orbits are based only on resolved measurements.

 In  3-body systems,  the osculating orbits  evolve with  time. In
 many   cases,  this  evolution   is  too   slow  to   be  detectable
 \citep[e.g.][]{Hei1996},   otherwise  a   more   complete  dynamical
  analysis should  be made, like  in \citep{Xu2015}. Such  analysis is
  beyond the scope of this paper. 

The   four   multiple  systems   studied   here   are  introduced   in
Table~\ref{tab:list}.  It  gives the  WDS codes and  names (discoverer
designations)   \citep{WDS},   the   {\it   Hipparcos}   numbers   and
trigonometric  parallaxes,  combined  visual magnitudes  and  spectral
types.  The  remaining columns list the  orbital parameters determined
here  (period  $P$  and  semimajor axis  $a$),  estimated  component's
masses, the  angles $\Phi$ between  the orbital momentum  vectors, and
the ratio  of outer and inner  periods $P_2 /P_1$.   The first system,
LHS~1070, is  composed of low-mass stars  or brown dwarfs  and has the
smallest ratio of $P_2 /P_1 = 4.5$, suggesting a 9:2 MMR.  The members
of the remaining three hierarchies have masses slightly below or above
solar.

The method of orbit calculation by simultaneous fitting of the inner
and outer pairs was presented by \citet{trip}; it is briefly outlined
in Section~\ref{sec:method}. Then in Section~\ref{sec:systems} the four
hierarchies are described individually. A few similar
systems discovered only recently are presented in Section~\ref{sec:new};
the paper concludes by the short summary in Section~\ref{sec:sum}.

\section{Calculation of triple-star orbits}
\label{sec:method}

\begin{deluxetable*}{c l rrr rrr r }
\tabletypesize{\scriptsize}
\tablewidth{0pt}
\tablecaption{Orbital Elements \label{tab:orb}}
\tablehead{
\colhead{WDS/system} &
\colhead{$P$} & 
\colhead{$T  $} &
\colhead{$e$} & 
\colhead{$a$} & 
\colhead{$\Omega$} &
\colhead{$\omega$} &
\colhead{$i$}  &
\colhead{$f$}    \\
\colhead{Name} &   
\colhead{(yr)} & 
\colhead{(yr)} &
\colhead{ } & 
\colhead{($''$)} & 
\colhead{(\degr)} &
\colhead{(\degr)} &
\colhead{(\degr)} 
}
\startdata
00247$-$2653/inner  &     17.247        &  2006.440 & 0.0172 &   0.4598 &    14.82     &   202.53 &    62.04 &     $-$0.485\\
LEI 1 BC          & $\pm$0.016 & $\pm$0.007 & $\pm$0.0008  & $\pm$0.0007 & $\pm$0.12 & fixed     & $\pm$0.11  & $\pm$0.006\\
00247$-$2653/outer &     77.62    &  2049.67   &   0.039  &   1.528         &    13.9 &   210.7 &    62.5 &        \\
LEI 1 A,BC        & $\pm$2.10    & $\pm$1.32   & $\pm$0.02 1 & $\pm$0.112 & $\pm$0.7 & $\pm$6.4 & $\pm$0.4 & \\
02022$-$2402/inner &     14.40 &  2006.400  &   0.058    &   0.102 &   120.5     &   227.2     &   152.5 &      $-$0.473\\
TOK 41 Ba,Bb      & $\pm$0.89 & $\pm$0.58 & $\pm$0.051 & $\pm$0.005 & $\pm$13.2 & $\pm$ 16.9 & $\pm$4.3 &  $\pm$   0.017\\
02022$-$2402/outer &     80.1   &  2041.75 &   0.336     &   0.414    &   179.5  &     0   &   180    &      \\
HDS 272 AB        & $\pm$ 6.1 & $\pm$2.25 & $\pm$0.043 & $\pm$0.038 & $\pm$3.3 &  fixed &  fixed & \\
05239$-$0052/inner &     47.03 &  1959.88   &   0.288&   0.3366      &   321.03 &   277.7   &    87.60 &     $-$0.592\\
A 847 B,C          & $\pm$0.36 & $\pm$0.40 & $\pm$0.014 & $\pm$0.0013 & $\pm$0.15& $\pm$0.80 &  $\pm$0.25     & $\pm$   0.016\\
05239$-$0052/outer &   1200  &  2280   &   0.20 &   3.716 &   335.41     &   249.2 &    95.67 &   \\
WNC 2 A,BC        & fixed  & $\pm$35 & fixed  & $\pm$0.020 & $\pm$0.45 & $\pm$7.8 &   $\pm$0.32  & \\
16057$-$3252/inner &     10.46  &  2012.623  &   0.240    &   0.1284    &   179.1  &    34.9  &   141.8 &       $-$0.491\\
WSI 84 Ba,Bb      & $\pm$0.07 & $\pm$0.027 & $\pm$0.006 & $\pm$0.0009 & $\pm$1.4 & $\pm$1.7 & $\pm$ 0.8 &   $\pm$0.008\\
16057$-$3252/outer &    131.2 &  1981.4 &   0.029      &   0.807 &   161.7 &    53.3 &   152.3 &         \\
SEE 264 A,B       & $\pm$1.9 & $\pm$4.0 & $\pm$0.013 & $\pm$0.070 & $\pm$3.8 & $\pm$12.6 & $\pm$1.4 &  

\enddata
\end{deluxetable*}

\begin{deluxetable*}{ll   r  rrr   rr c }
\tabletypesize{\scriptsize}     
\tablecaption{Relative positions and residuals (fragment)
\label{tab:pos} }  
\tablewidth{0pt}                                   
\tablehead{                                                                     
\colhead{WDS} & 
\colhead{Sys} & 
\colhead{Date} & 
\colhead{$\theta$} & 
\colhead{$\rho$} & 
\colhead{$\sigma$} & 
\colhead{O$-$C$_\theta$} & 
\colhead{O$-$C$_\rho$} & 
\colhead{Ref\tablenotemark{a}} \\
&   &  
\colhead{(year)} & 
\colhead{(\degr)} & 
\colhead{(\arcsec)} &
\colhead{(\arcsec)} &
\colhead{(\degr)} &
\colhead{(\arcsec)} &
}
\startdata
02022$-$2402 & Ba,Bb &  2008.6990 &    187.9 &   0.0880 &   0.0050 &      0.1 &  $-$0.0011 & s  \\
02022$-$2402 & Ba,Bb &  2008.7670 &    183.9 &   0.0830 &   0.0050 &     $-$2.0 &  $-$0.0065 & S \\
02022$-$2402 & Ba,Bb &  2008.7670 &    186.5 &   0.0914 &   0.0050 &     0.6    &    0.0019 & S  \\ 
02022$-$2402 & Ba,Bb &  2009.6700 &    162.2 &   0.1018 &   0.0050 &      0.5 &   0.0060 & S \\
02022$-$2402 & A,Ba &  2008.7674 &    343.7 &   0.5814 &   0.0050 &     $-$0.5 &   0.0008 & S \\
02022$-$2402 & A,Ba &  2009.6709 &    340.4 &   0.5851 &   0.0050 &     $-$0.0 &   0.0016 & S \\
02022$-$2402 & A,B &  1991.2500 &     28.0 &   0.5720 &   0.0100 &      3.0 &   0.0447 & H \\
02022$-$2402 & A,B &  2000.7670 &      0.6 &   0.5350 &   0.0050 &     $-$1.1 &  $-$0.0183 & s 
\enddata
\tablenotetext{a}{References: G: Gaia; H: Hipparcos, K: \citet{Koh2012}; M: micrometer measures; 
P: photographic; S: speckle interferometry at SOAR, s: other speckle interferometry. }
\end{deluxetable*}


The objects for this study  were selected among triple systems with one
(or  both)   orbits  already  listed   in  the  Sixth   orbit  catalog
\citep{VB6}.   The available  astrometry  was extracted  from the  WDS
database by  B.~Mason and complemented by  recent speckle measurements
at the Southern Astrophysical Research (SOAR) 4.1 m telescope, some of
those still unpublished.  One of the triples (HIP~9497) was discovered
at SOAR in 2008, another (HIP  78842) was discovered the same year and
with   the   same  speckle   instrument   at   the  Blanco   telescope
\citep{TMH10}.

I use the IDL code {\tt orbit3.pro} to fit simultaneously the position
measurements of the  inner and outer systems by  two Keplerian orbits;
see \citet{trip}  for further information. The two  orbits are defined
by 20 orbital elements.  However,  as for the objects studied here the
RVs are  not available,  only 15 ``visual''  elements are  fitted.  In
this paper, the  parameters of the inner orbit  (Ba,Bb) are denoted by
the index 1,  while the outer orbit (A,B) has the  index 2.  The inner
and  outer orbits  can be  computed independently  of each  other, but
then, to compute the outer orbit,  the positions of the center of mass
of  the inner subsystem  must be  calculated using  its orbit  and the
estimated component's masses.  Simultaneous fitting avoids assumptions
about  masses and  allows  measurement  of the  inner  mass ratio,  as
explained below.

Let A,B be  the outer pair and Ba,Bb the  inner subsystem belonging to
the secondary component.  The positions  of A,Ba reflect the motion in
both  orbits, resulting  in the  wavy trajectory  (wobble).   When the
inner  pair is  not resolved,  the measurements  of A,B  refer  to the
photo-center of  the subsystem  Ba,Bb and may  still show a  wobble of
lower amplitude.  The  ratio of the wobble amplitude  to the semimajor
axis  of the  inner  orbit, called  here  {\it wobble  factor}, is  $f
=q_1/(1+q_1)$ for  resolved measurements and $f^* =q_1/(1+q_1)  - r_1/(1+r_1)$ for
the  photo-center measurements,  where $q_1$  is the   mass ratio  in the
inner  pair  and  $r_1$ is  the  light  ratio  of its  components.   Let
${\mathbf x}_{\rm  A,B}$ be  the position vector  in the  outer orbit,
directed from A to the center of mass B, and ${\mathbf x}_{\rm Ba,Bb}$
the vector  in the  inner subsystem. Then  the vector of  the resolved
position in the outer pair is
\begin{equation}
{\mathbf x}_{\rm A,Ba} = {\mathbf x}_{\rm A,B}  + f \; {\mathbf  x}_{\rm
  Ba,Bb} .
\label{eq:wobble}
\end{equation}

The IDL code  was modified here to treat  both resolved and unresolved
measurements of the outer  pair ({\tt orbit3.pro} allows only resolved
measurements).   If   the  inner  subsystem  is  a   twin  with  equal
components, $f^* =  0$, and unresolved measurements of  the outer pair
do  not  contain  any  wobble.   When the  subsystem  belongs  to  the
secondary component,  the wobble factor is negative,  but the relation
$q_1 = |f|/(1 - |f|)$ still holds.

The angle $\Phi$ between the angular momentum vectors of the inner and
outer  orbits (mutual  inclination) can  be easily  computed  from the
orbital elements $\Omega$ and $i$. However, without RV information the
true ascending  nodes of both orbits remain  undefined; an alternative
orbit with  a pole reflected around  the line of  sight corresponds to
the same on-sky measurements.  Two alternative values of $\Phi$ result
from this situation, and it is not known {\it a priori} which angle is
the correct one.  A smaller  angle is normally selected on statistical
grounds, considering the orbit alignment tendency \citep{moments}, and
by using other  arguments such as growth of  the inner eccentricity by
Kozai-Lidov cycles at large  $\Phi$ \citep{Naoz2016}.  The last column
of Table~\ref{tab:list}  gives both alternative angles  $\Phi$ and the
period ratio $P_2/P_1$.

The  orbital elements  derived here  and  their errors  are listed  in
Table~\ref{tab:orb}, in common notation. For the inner subsystems, the
last  column gives  the wobble  factor $f$.   The  individual position
measurements and their  residuals are provided in Table~\ref{tab:pos},
available  in full  electronically.  Its  second column  specifies the
nature of  the measurement: Ba,Bb  refers to the resolved  inner pair,
A,Ba to the resolved outer pair,  and A,B to the unresolved measure of
the outer pair.   As the errors of positional  measures are either not
provided in the original data sources or are unreliable, the errors in
Table~\ref{tab:pos}  are  assigned   subjectively,  depending  on  the
observing  technique, to  determine  the relative  weights.   The
  errors   of  modern   measures  by   adaptive  optics   and  speckle
  interferometry at  large telescopes  are between 2\,mas  and 5\,mas,
  the {\it Hipparcos} relative  positions are accurate to 10\,mas, the
  photographic  astrometry has errors  of 30\,mas,  and the  errors of
  visual  micrometer  measures   range  from  50\,mas  to  0\farcs25.
Outlying visual  measures are given artificially low  weight to reduce
their  impact.  The actual  weighted rms  residuals roughly  match the
adopted errors. 

In  the  following  Figures,  I  plot the  outer  trajectory  (or  its
fragment) with the  wobble included.  One outer period  is plotted, so
the trajectory is  not closed when the period ratio  is not an integer
number.   Resolved measures  are plotted  as large  asterisks  and are
connected  to the  ephemeris  positions by  short  dotted lines.   The
unresolved  measures of  the  outer pair  are  plotted by  smaller
symbols and also connected  to their ephemeris positions corresponding
to the smaller  wobble factor $f^*$; these positions do  not lie on the
same trajectory.  The inner orbit  and measures are over-plotted by the
dashed line  and triangles  around the same  center, even  though here
they depict the motion of the {\it secondary} subsystem. In all plots
the scale is in arcseconds, North is up,  East is left.

\section{Individual systems}
\label{sec:systems}

The  four hierarchical  systems featured  in  Table~\ref{tab:list} are
discussed in the following sub-sections.

\subsection{00247$-$2653 (LHS 1070)}

\begin{figure}[h]
\epsscale{1.0}
\plotone{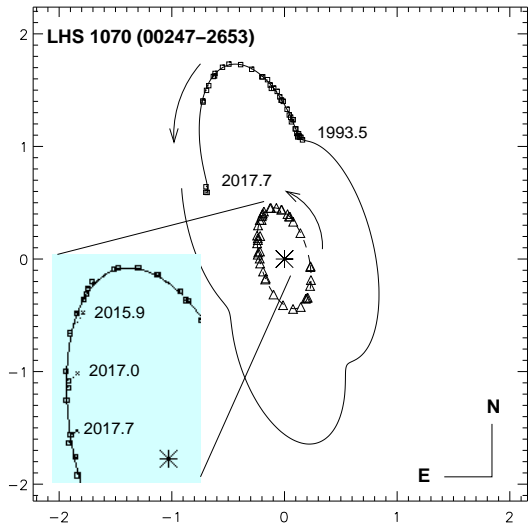}
\caption{The  orbits of  LHS  1070 (WDS  J00247$-$2653).  In this  and
  subsequent plots, the axis scale is in arcseconds. The insert shows
  fragment of the inner orbit where the deviating measures made
  in 2015--2017 are plotted by crosses.
\label{fig:LHS1070}  }
\end{figure}

The triple system LHS~1070, also  known as GJ~2005, LP 881-64, and WDS
J00247$-$2653,  is  located  at  7.7\,pc  from the  Sun.   The  recent
parallax  of 132.3$\pm$11.4\,mas  given  by \citet{Weinberger2016}  is
less accurate  than 129.5$\pm$2.5\,mas measured  by \citet{Costa2005},
so the latter  value is used here.  The triple  system consists of two
stars  B and  C with  masses around  0.07 ${\cal  M}_\odot$  (near the
hydrogen burning limit) in a 17.25 year orbit around each other.  They
are accompanied by  the more distant and massive  primary component A.
Detailed  analysis of  this  hierarchy, including the review of  relevant
literature,  was  presented by  \citet{Koh2012}.   These authors  have
shown that the system is dynamically stable only when the outer period
exceeds  $\sim$80 years  and the  outer eccentricity  is  small (their
Fig.~5). Yet, they preferred the {\it unstable} outer orbit of 44 year
period that provided  the best fit to the data  available at the time,
covering the  period from 1993 to  2008.  This orbit  still figures in
the  catalog  \citep{VB6},   challenging  common  dynamical  stability
criteria.   New measurements  made at  SOAR  in 2015--2017  lead to  a
different outer orbit and resolve the discrepancy.

The  revised   orbital  elements  and  their  errors   are  listed  in
Table~\ref{tab:orb}. I  have chosen to fix the  inner angle $\omega_1$
in  the  final  fit  to  avoid  a large  error  of  $T_1$  (the  inner
eccentricity is small and  $\omega_1$ strongly correlates with $T_1$).
The wobble  factor $f  = 0.485  \pm 0.006$ means  $q_1 =  f/(1 -  f) =
0.942$.

The  inner orbit  computed here  is essentially  identical to  that by
\citet{Koh2012}, as it already has  been defined with a high precision
by   prior   measurements.    The   parallax   of   129.5$\pm$2.5\,mas
\citep{Costa2005} leads  to the  inner mass sum  of 0.150  $\pm$ 0.009
${\cal  M}_\odot$, hence  the  individual masses  of  0.077 and  0.070
${\cal M}_\odot$.  Note,  however, that the latest measures  of B,C at
SOAR have  large and  similar residuals to  the new  orbit ($-6\degr$)
that cannot be removed by  its adjustment.  This is most likely caused
by  the real  deviations of  the  inner subsystem  from the  Keplerian
motion  owing to  its  interaction  with the  tertiary.    A  full
  dynamical analysis of this system  is in order. Meanwhile, the SOAR
measures of B,C were given reduced weight in the orbit fit. 

The  outer  orbit converged  to  a  solution  that is  almost  exactly
coplanar  with the inner  orbit.  The  period ratio  is 4.50$\pm$0.03.
Note that  the inner and outer  lines of apsides  are aligned (similar
$\omega$). Figure~\ref{fig:LHS1070} shows both  orbits. The outer mass sum
of 0.273  ${\cal M}_\odot$  leads, by subtraction,  to the mass  of A,
0.123 ${\cal M}_\odot$.

The  period ratio  of  4.5 is  slightly  less than  the 4.7  canonical
stability limit  derived by \citet{MA2001} for  coplanar triples.  The
system is  thus on the  verge of dynamical stability,  implying strong
dynamical interaction between the orbits.  It is very likely that this
interaction drives the periods into a MMR. The period ratio found here
implies the 9:2 MMR.

The orbit coplanarity strongly suggests viscous evolution in a disk as
the  formation mechanism  of this  triple system.   The  pair B,C
formed first and migrated to the present-day separation of 3.5 au as a
result  of  accretion,  until  the  outer  companion  A  overtook  the
accretion flow.  The orbit of  A also evolved to a smaller separation,
reaching  its present  axis  of  12 au.   If  the migration  proceeded
further, the  triple system would have  broken up into a  binary and a
single  star.  The  fact that   migration  stopped just  before the
break-up   explains  the  rarity  of  similar  quasi-stable  triple
systems. 

\citet{Koh2012} noted  that the component  A is $\sim$3 mag  too faint
for its mass (or over-massive for its luminosity). They suggested that
A could be itself a close binary, but have refuted its resolution into
a closer  pair, reported previously. The  new outer orbit  leads to a
slightly smaller mass of A, but does not resolve the mass discrepancy.

\subsection{02022$-$2402 (HIP~9497)}

\begin{figure}
\epsscale{1.0}
\plotone{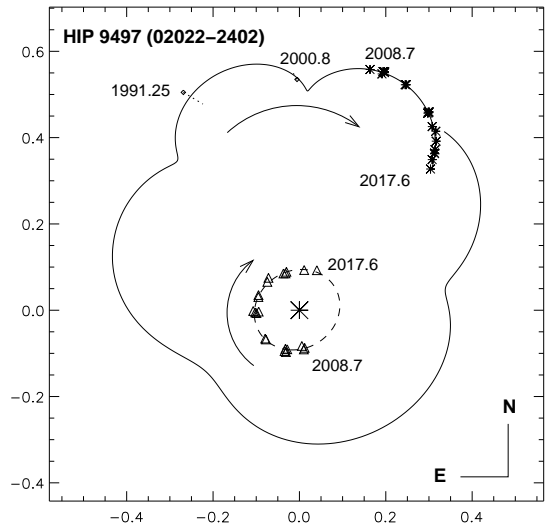}
\caption{The orbits  of HIP~9497 (WDS  J02022$-$2402, HDS 272  A,B and
  TOK 41 Ba,Bb). 
\label{fig:HIP9497}  }
\end{figure}

The   triple   system  HIP~9497   was   discovered   in  2008   nearly
simultaneously  and  independently  at  SOAR  \citep{TMH10}  and  WIYN
\citep{Horch2012}   telescopes  by   resolving  the   faint  secondary
component  of the  {\it Hipparcos}  binary HDS~272  into a  close pair
TOK~41 Ba,Bb.  Now about 60\%  of the inner  orbit is covered  and its
elements are well constrained.  However, in 26 years elapsed since the
discovery of the outer pair A,B, only a third of its orbit is covered,
so its elements are still  uncertain.  Here the previous tentative 138
year orbit of A,B is updated to  $ P_2 = 80$ years.  I reprocessed the
published SOAR  measure of 2008.76 where  Ba and Bb  were swapped.  As
the outer  orbit is  seen almost face-on,  I fixed its  inclination to
$i_2 = 180\degr$, which makes the two angles $\Omega_2$ and $\omega_2$
degenerate; hence, $\omega_2  =0$ is also fixed.  The  period ratio is
obviously small,  $P_2/P_1 = 5.6 \pm  0.7$. The still  low accuracy of
the period  ratio does not  allow to make  any statements on  the MMR,
although the  resonance is likely.  The resolved  measurements of A,Ba
clearly   show   the  wavy   motion   caused   by   the  inner   orbit
(Figure~\ref{fig:HIP9497}).

The {\it  Hipparcos} parallax  of 16.09$\pm$0.76\,mas is  adopted here
because the {\it Gaia} \citep{Gaia} parallax of 17.31$\pm$0.99\,mas is
less  accurate. The  parallax and  the inner  orbit correspond  to the
inner mass sum of 1.22$\pm$0.17  ${\cal M}_\odot$, hence the masses of
0.64 and  0.58 for  Ba and  Bb, respectively (the  wobble factor  $f =
-0.473 \pm 0.020$  implies the inner mass ratio  of 0.90).  They agree
with  the masses  estimated  from the  absolute  magnitudes   (the
  interstellar  extinction is negligible  for these  nearby systems).
The less well defined outer orbit gives the 1.44 ${\cal M}_\odot$ mass
of A,  matching its spectral type  F6V.  This triple  system, like the
previous one, is a double twin.

The inner orbit is almost circular  ($e_1 = 0.06 \pm 0.05$), while the
outer eccentricity $ e_2 = 0.33 \pm  0.04 $ is moderate. As one of the
orbits  has a  face-on  orientation, the  two alternative  inclination
angles coincide,  $\Phi = 27$\degr. At such  mutual inclination, there
are no Kozai-Lidov cycles.

\subsection{05239$-$0052 (HIP 25240)}

\begin{figure}
\epsscale{1.0}
\plotone{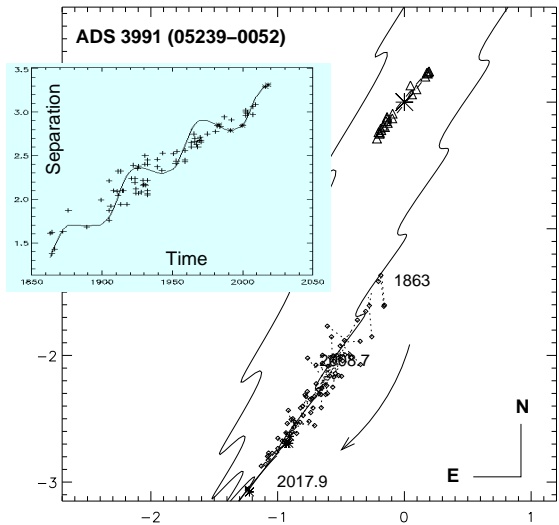}
\caption{Orbits of ADS 3991 (WDS J05239$-$0052). The insert shows the
  separation between A and BC (plus signs) and A and B (asterisks)
  vs. time, the wavy line is the modeled separation of A and B. 
\label{fig:ADS3991}  }
\end{figure}

This is a  bright 2+2 quadruple system HR~1782  (ADS 3991, HIP 25240).
The  double-lined spectroscopic  pair Aa,Ab  has a  period  of 22\fd58
\citep{Tok1997}, while  the visual pair B,C has  a well-defined visual
orbit with $P_1  =47$ years.  The two inner  pairs revolve around each
other on a  wide, poorly constrained outer orbit  with $P_2 \sim 1000$
years.  Given  this freedom, I  chose to fix  the outer period  $P_2 =
1200$ years  and the  outer eccentricity $e_2  = 0.2$ in  the combined
orbit fit, which gives then a reasonable outer mass sum. The published
measure of the 3\arcsec A,B pair  made at SOAR in 2015.9 was distorted
by aliasing; here  it is corrected and confirmed  by the fresh measure
in 2017.9.   The last unresolved measure  of A,BC is  provided by {\it
  Gaia}.

Both inner and outer orbits are highly inclined and nearly parallel on
the sky (Figure~\ref{fig:ADS3991}).   The wobble is therefore parallel
to the  outer trajectory and  affects mostly the  separation.  Indeed,
the five  resolved measures of A,B  made from 1983 to  2017 by speckle
interferometry and {\it Hipparcos} demonstrate the 47-year wave in the
separation and define the wobble  factor $f = -0.592 \pm 0.016$.  This
means  that the  component C  is more  massive than  B, $q_{\rm  B,C} =
1.44$. 

The {\it Hipparcos}  parallax of $17.95 \pm 0.77$  mas could be biased
by  the complex  nature of  this source  consisting of  three resolved
stars.  The {\it  Gaia} parallax of 18.75$\pm$0.47 mas  results in the
inner mass sum of B,C of 2.62 ${\cal M}_\odot$, larger than 2.3 ${\cal
  M}_\odot$ derived  from the absolute  magnitudes in \citep{Tok1997}.
Considering  the  large  wobble  amplitude,  it  is  likely  that  the
subsystem  B,C actually contains  an additional  close companion  to C
with a mass of $\sim$0.5 ${\cal M}_\odot$; the derived masses of B and
C are  1.07 and 1.55 ${\cal  M}_\odot$, respectively. The  mass sum in
the outer orbit is 5.4 ${\cal M}_\odot$.

The RVs of A (center of mass of Aa,Ab) and BC, measured around 1994 by
\citet{Tok1997},  were 55.0$\pm$0.2 and  52.1$\pm$0.2 km~s$^{-1}$,
respectively.  The  RV difference  between A and  BC could be  used to
further constrain the outer orbit  by requiring that the RV amplitudes 
match the  expected mass  sum.  This can  be achieved by  imposing the
constraint $\omega_2  = 300\degr$,  leading to a  shorter $P_2$  and a
larger $e_2$.  However, the measured  RV of BC (combined light of both
components) could be  biased by the orbital motion in  this pair, so I
do not trust its value and only assume that the sign of the difference
is  correct and  thus defines  the true  ascending node  of  the outer
orbit.

Although  the outer orbit  is poorly  constrained, its  nearly edge-on
orientation  allows  a robust  estimate  of  the relative  inclination
between the orbits of A,BC and B,C: $\Phi = 16\fdg5$. As the ascending
node of B,C  is not known, the alternative angle  $\Phi = 165\degr$ is
also possible, although less likely.


\subsection{16057$-$3252 (HIP 78842)}

\begin{figure}
\epsscale{1.0}
\plotone{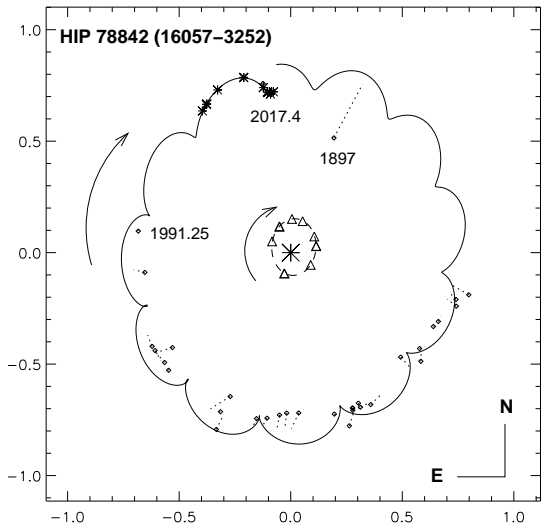}
\caption{ The orbits  of HIP~78842 (WDS J16057$-$3252, SEE  264 A,B and
  WSI 84 Ba,Bb).
\label{fig:SEE264}  }
\end{figure}

The   hierarchical   system   HIP~78842  ({\it   Hipparcos}   parallax
24.70$\pm$1.96 mas) is  fortunate in the coverage of  both its orbits.
The outer  pair SEE~264 A,B, discovered  in 1897, has  made almost one
full revolution,  defining its 131 year nearly  circular face-on orbit
quite well. The  secondary component B was resolved  into a close pair
WSI~84 Ba,Bb  in 2008.55 \citep{TMH10}  and has also  completed almost
one full revolution; its 10 year  orbit is computed here for the first
time.  The  rms residuals of the  11 speckle measurements  of Ba,Bb to
this orbit are remarkably small, only 1.4\,mas.

Figure~\ref{fig:SEE264} illustrates the  ``dancing'' orbital motion in
this  triple system.   The two  orbits  are nearly  coplanar, $\Phi  =
14$\degr  (the alternative  inclination $\Phi  = 65\degr$  would cause
Kozai-Lidov  cycles  and  hence  is  unlikely). The  period  ratio  of
12.6$\pm$0.2  is small,  while  the MMR  cannot  be ruled  out at  the
current accuracy.  The  wobble factor $f = -0.491  \pm 0.008$ leads to
$q_1 =  0.96$, confirming  that the inner  subsystem Ba,Bb is  a twin.
The unresolved  historic measurements of  A,B do not show  any wobble,
and this  is taken into account  in the orbital fit  (for this reason,
most short  dotted lines in  Figure~\ref{fig:SEE264} do not  touch the
outer   trajectory  depicted   by  the   full  line). The   measure  by
\citet{Msn2011d} was corrected because Ba and Bb were swapped and the
outer measure refers to A,Bb, not to A,B as published. The last unresolved
measure of A,B is furnished by {\it Gaia}.

Considering the large error of  the {\it Hipparcos} parallax, I prefer
to  use the  dynamical parallax  of 24.6\,mas  derived from  the inner
orbit by adopting  the masses of 0.75 ${\cal M}_\odot$  for Ba and Bb,
estimated from their absolute magnitudes. The mass of A derived in the
same  way is  0.96 ${\cal  M}_\odot$.  The  less accurate  outer orbit
gives  the matching  dynamical  parallax of  23.3\,mas.  

Unlike  other hierarchies  studied  here, HIP~78842  is  not a  double
twin. It is  actually a 3+1 quadruple, considering  the companion C at
9\farcs3 separation  with matching proper motion,  RV, and photometric
distance; the estimated period of  AB,C is $\sim$4 kyr.  The estimated
mass of C,  0.67 ${\cal M}_\odot$, is similar to the  masses of Ba and
Bb; its spectral type is K5V.   The combined color of AB is bluer than
that  of  C:   their  $V-K$  indices  are  2.46   mag  and  3.32  mag,
respectively.

The  quasi-circular  and aligned  orbits  of  A,B  and Ba,Bb  strongly
suggest  that  this  hierarchy  was  formed in  a  rotating  disk,  as
discussed in Section~\ref{sec:intro}.  Moreover, the circularity of the A,B
orbit implies  the absence of  Kozai-Lidov cycled caused by  the outer
companion  C, so  its  orbit  should also  be  inclined by  $<39\degr$
relative to the orbit of A,B. The presence of C fits the sequential
formation scenario and possibly explains why the component A could not
grow  to  become  a double  twin:  the  gas  supply  coming to  A  was
re-directed  to the  distant component  C when  it  formed.  Something
similar may  have happened in  the ``planetary'' 3+1  quadruple system
HD~91962  \citep{planetary}, where all  three secondary  components are
less massive than the primary star.

\begin{figure}[ht]
\epsscale{1.0}
\plotone{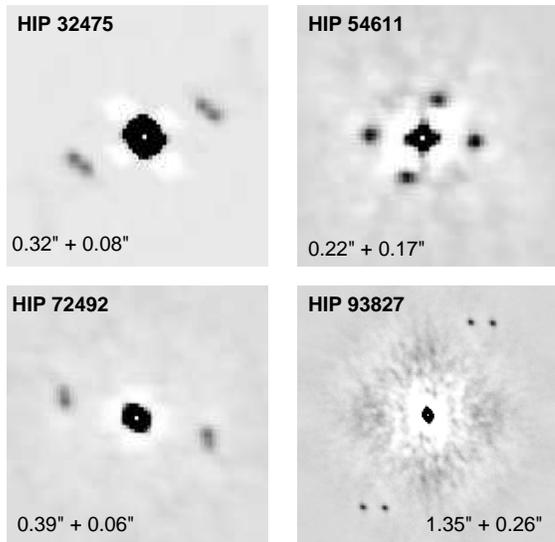}
\caption{Double twins discovered at SOAR in 2015 and 2016. Each panel
  shows the speckle auto-correlation function in arbitrary scale and
  in negative intensity stretch. The Hipparcos numbers and separations
  are indicated.
\label{fig:mosaic}  }
\end{figure}

\section{Recently discovered double twins}
\label{sec:new}

\begin{deluxetable}{c c c  cc  }
\tabletypesize{\scriptsize}
\tablewidth{0pt}
\tablecaption{Double twins discovered at SOAR
\label{tab:new} }
\tablehead{
\colhead{WDS} &
\colhead{HIP} & 
\colhead{$M_1$} &
\colhead{$P^*_{\rm out}$} & 
\colhead{$P^*_{\rm in}$} \\ 
&  &
\colhead{(${\cal M}_\odot$)} &
\colhead{ (yr)} &  
\colhead{ (yr)}   
}
\startdata
06467+0822   & 32475 & 1.45 & 64  & 11  \\
11106$-$3234 & 54611 & 1.96 & 900 & 113 \\
14494$-$5726 & 72492 & 1.59 & 190 & 19  \\
19064$-$1154 & 93827 & 1.11 & 200 & 31  
\enddata
\end{deluxetable}

The luminosity of dwarf stars is a strong function of their mass. In a
double twin  triple system, the  combined light of  Ba and Bb  is much
fainter  than  the  light  of  A.  Binaries  with  a  large  magnitude
difference  $\Delta m$  are difficult  to discover.  Consequently, the
number of double twins among  low-mass stars in the solar neighborhood
could be substantially larger than known currently.

New  observing   techniques  such  as  adaptive   optics  and  speckle
interferometry at  large telescopes help  to discover double  twins by
resolving secondary components of  known binaries with a large $\Delta
m$  into inner  pairs.  Four  such discoveries  made at  SOAR  in 2015
\citep{Tok2016}    and    2016    \citep{Tok2018}   are    shown    in
Table~\ref{tab:new}  and illustrated in  Figure~\ref{fig:mosaic}.  All
are previously  known binaries  with primary components  slightly more
massive than the  Sun, where the faint secondary  components have been
resolved  at SOAR  into  close  pairs composed  of  equal stars.   The
orbital periods  $P^*$, estimated  crudelyfrom the  separations, indicate that
after several  decades of monitoring  the inner and outer  orbits will
become defined  and the architecture of these  hierarchies will become
known.   The system  HIP~54611 looks  non-hierarchical.   However, the
outer pair was discovered in  1879 at a larger 1\farcs0 separation and
has  closed  down  since;  apparently,  the inner  subsystem  B,C  now
projects on the primary component A.

The  latest version  of  the Multiple  Star  Catalog, MSC  \citep{MSC}
contains   many   other  double twins  and 2+2  quadruples; their
orbital  periods  range  from  days  to  kilo-years.   Some  of  those
hierarchies  are composed  of low-mass  stars.   To give  just two
  examples, WDS  J04185+2817 consists of two resolved  pairs of M3.5V
dwarfs,  DD~Tau and  SZ~Tau, separated  by 30\arcsec  (estimated outer
period 3.8  kyr).  The  WDS J05101$-$2341 is  also a 2+2  quadruple of
similar architecture with M3V components.

\section{Summary and discussion}
\label{sec:sum}

\begin{figure}
\epsscale{1.0}
\plotone{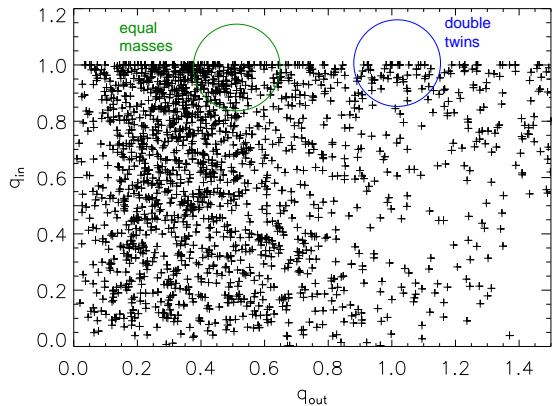}
\caption{Comparison of mass ratios at adjacent hierarchical levels for
  multiple  systems  from the  MSC  with  primary  mass $<$1.5  ${\cal
    M}_\odot$.
\label{fig:qlqs}  }
\end{figure}

The four  hierarchical systems studied  here have several  features in
common:  moderate ratios  of  inner and  outer periods,  approximately
co-aligned orbits  with small eccentricity, and similar  masses in the
inner   and  outer   subsystems  (double   twins).   As   outlined  in
Section~\ref{sec:intro}, such  hierarchies could be  produced by sequential
formation  and migration  of their  components  at the  epoch of  mass
assembly, i.e. at  the protostellar phase. It is  difficult to imagine
such well-organized systems to be products of chaotic dynamics.

Sequential formation may not always produce double twins. If the outer
companion forms by the end of  the accretion phase, it may not grow in
mass sufficiently to equal  the inner pair.  Alternatively, the growth
of the  outer companion can be  stopped by formation  of another, more
distant star, as might have happened in the case of HIP~78842.

The triple system LHS~1070 is remarkable by its low total mass of 0.27
${\cal  M}_\odot$.  However,  it  does  not hold  the  record in  this
respect. The two lowest-mass triples in the MSC, WDS J02055$-$1159 and
J08382+1511, have the estimated total  masses of 0.150 and 0.165 ${\cal
  M}_\odot$, respectively,  and are  composed of brown  dwarfs.  Their
architecture  is similar  to that  of  LHS~1070: they  consist of  the
primary component  A and the  secondary subsystem B,C with  a moderate
ratio  of separations  and the estimated  periods  of a  few decades  (the
orbits  are not  known  yet).  However,  neither  of these  sub-stellar
hierarchies are  double twins because the  mass of A is  less than the
total mass of the inner subsystem.

Figure~\ref{fig:qlqs}  compares  the   mass  ratios  at  two  adjacent
hierarchical levels computed for all systems from the MSC with primary
mass less than  1.5~${\cal M}_\odot$.  As this catalog  is burdened by
large  selection  effects, the  Figure  does  not represent  unbiased
statistics. However, if the mass-ratio distribution has sharp details,
they  could  probably  be  seen  in this  plot  because  observational
selection is expected  to be a smooth function  of parameters. Indeed,
concentration  of  the  inner   mass  ratios  toward  one  (twins)  is
evident. On the  other hand, the points do not  show any clustering at
the  (1,1) location  corresponding  to double  twins.  The density  of
points is  larger near  (0.5,1), i.e. when  all three  components have
comparable masses.   However, as noted above, the  discovery of double
twins may have been seriously hampered until now.

Determination of  inner and  outer orbits in  triple systems  is paced
by  the  accumulation  of  measurements. In  most  cases,  the
available time coverage does not yet allow meaningful analysis of long
outer  orbits.  The  work presented  here can  be expanded  to include
several  other hierarchies, but  their number  will remain  modest and
will  grow only slowly  with time.   On the  other hand,  discovery of
double twins  using modern  high-resolution techniques is  a promising
undertaking,  especially when  applied to   nearby  low-mass stars.
Systematic surveys  of such  volume-limited samples with  high angular
resolution are under way already \citep[e.g.][]{Law2010}.

\acknowledgements
Some data used here were obtained at the Southern Astrophysical
Research  (SOAR)   telescope. 
This work  used the  SIMBAD service operated  by Centre  des Donn\'ees
Stellaires  (Strasbourg, France),  bibliographic  references from  the
Astrophysics Data  System maintained  by SAO/NASA, and  the Washington
Double  Star  Catalog  maintained  at  USNO.   I  thank  B.~Mason  for
extracting historic measurements from the WDS database.

\facility{Facility: SOAR}

\end{document}